\documentclass[%
 reprint,
superscriptaddress,
nofootinbib,
 amsmath,amssymb,
 aps,
]{revtex4-2}

\usepackage{graphicx}
\usepackage{dcolumn}
\usepackage{bm}
\usepackage{hyperref}
\usepackage{braket}
\usepackage{mathrsfs}

\begin{document}

\preprint{APS/123-QED}

\title{Inflationary Quantum Spectrum of the Quasi-Isotropic Universe}

\author{Nicola Bortolotti}
 \email{nicola.bortolotti@uniroma1.it}
 \affiliation{Physics Department, ``Sapienza” University of Rome, P.le Aldo Moro 5, 00185 (Roma), Italy}
 \affiliation{Centro Ricerche Enrico Fermi, Via Panisperna 89a, 00184 Rome, Italy}
\author{Giovanni Montani}%
 \email{giovanni.montani@enea.it}
\affiliation{Physics Department, ``Sapienza” University of Rome, P.le Aldo Moro 5, 00185 (Roma), Italy}
\affiliation{ENEA, FSN-FUSPHY-TSM, R.C. Frascati, Via E. Fermi 45, 00044 Frascati, Italy
}%

\date{\today}

\begin{abstract}
We investigate the quantum dynamics of the quasi-isotropic inflationary solution. This is achieved by deriving the Lagrangian and Hamiltonian for both the FLRW background and the inhomogeneous correction, via an expansion of the Einstein-Hilbert action up to second order in the perturbation amplitudes. We do this in a gauge invariant fashion adopting the Mukhanov-Sasaki variable for the scalar degree of freedom. Then we construct the quantum dynamics in terms of a semiclassical WKB scenario for which the inhomogeneous component of the Universe is treated as a ``small" quantum subsystem, evolving on the classical isotropic background. Starting from the Wheeler–DeWitt equation, we recover a Schr\"odinger dynamics for the perturbations, in which the time dependence of the wave function emerges thanks to the classicality of the background, and we solve it for a de Sitter inflationary phase. The main result of this paper is that the resulting scalar component of the power spectrum has the standard scale invariant profile, while the tensor one results to be not constrained by the inflationary expansion (modulo an overall normalization factor), therefore preserving the spatial distribution of the quasi-isotropic correction to the metric.
\end{abstract}


\maketitle

\section{Introduction}

The isotropic Universe, as described by the Robertson-Walker geometry \cite{robertson1935kinematics,walker1936,montani2011primordial}, models a large part of the present Universe evolution, as confirmed by the nucleosynthesis of the light elements \cite{kolb2018early} and, overall, the isotropy of the microwave background temperature \cite{planck2020isotropy}. 

However, as outlined mainly by the activity of the Landau School, the Robertson-Walker geometry is far from being the most general dynamical approach to the cosmological singularity \cite{belinskii1982general,montani2011primordial,Gravitation,misner1968isotropy,MontaniReview}. 

An interesting generalization of the isotropic Universe has been derived in \cite{LKoriginal} by Lifshitz and Khalatnikov, where, by a Taylor expansion in the synchronous time of the metric component, it was constructed the so-called \textit{quasi-isotropic solution}. Its physical meaning is given by the fact that it represents the most generic spatially inhomogeneous generalization of the FLRW model in which the spacetime is locally FLRW-like near the cosmological singularity. This solution, existing in the presence of ultrarelativistic matter, contains three physically arbitrary functions of the spatial coordinates and the metric splits into a background isotropic-like component and a small correction, controlled by a different scale factor with respect to the background one. This solution has a certain degree of isotropy, although being an inhomogeneous cosmology. For other generalization of the quasi-isotropic solution in the presence of a scalar field, of viscous dissipation and during the de Sitter phase, see \cite{montani1999quasi,montani2008QIviscous,montani2003quasi} respectively. It is worth remarking that, for such different physical contexts, the quasi-isotropic solution is no longer a Taylor expansion of the three-metric in the synchronous time, but it retains the same structure with two (one larger and one smaller) scale factor, having, in principle, generic functional form, as result of the dynamics - for instance, in the presence of a scalar field these scale factors are suitable power laws. Here, we study the quasi-isotropic solution during a de Sitter phase of the Universe and in the presence of an inhomogeneous quantum scalar field. In other words, we are considering the generation of the primordial inflaton and gravitational field fluctuations during the inflationary expansion \cite{weinberg2008cosmology,kolb2018early}. 

\vspace{5mm}

The peculiar point of our analysis consists of treating the correction, due to the small scale factor (actually, the ratio of the smaller to the larger scale factor), as a quantum degree of freedom.

We implement to our model the ideas discussed in \cite{vilenkin1989interpretation,kiefer1991quantum}, see also \cite{maniccia2022wkb}. That is, we treat the inhomogeneous correction to the metric and the scalar field as a ``small'' quantum subset, living on the classical isotropic background. Thus, from a WKB expansion in the Planck constant of the Wheeler-DeWitt (WDW) equation and implementing a Born-Oppenheimer-like approximation, we recover a Schr\"odinger-like dynamics for the two quantum degrees of freedom, where the time dependence of the wave function comes out from the dependence of the background metric on the label time (for a critical analysis and a reformulation of this scheme, when the WKB expansion is extended to higher orders, see \cite{maniccia2021nonunitarity,maniccia2022quantum}). 

After a detailed construction of the Lagrangian and Hamiltonian for the model, as splitted into the zero, first and second order of approximation in the perturbation variables, we derive the quantum dynamics of the metric (actually a tensor degree of freedom) and the scalar field. To deal with a gauge invariant formulation, we adopt the Mukhanov-Sasaki variable for the scalar component, and we retain as factorized between a time dependent and a space dependent function the tensor degree of freedom. In other words, preserving also on a quantum level the structure of the quasi-isotropic solution has the non-trivial implication that the spatial dependence of the metric is not coupled to the quantum evolution of the tensor component. The scalar field is, instead, treated as an intrinsically inhomogeneous degree of freedom and it is analyzed via a Fourier decomposition, leading to a mini-superspace dynamics for each independent mode. 

\vspace{2mm}

The main result of the present study is to outline the real nature of the primordial spectrum, associated to the proposed dynamical scenario. In particular, while the scalar field, i.e. the scalar component of the spectrum, corresponds to the usual scale invariant profile of the predicted spectrum, the tensor component preserves the generic fingerprint contained in the space dependence of the quasi-isotropic correction, and we show that such a contribution is smaller than the scalar one (the tensor to scalar ratio remains much less than unity), but it is not suppressed. 

The merit of the present analysis is to demonstrate that, under suitable conditions, corresponding to the validity of the quasi-isotropic scheme also in the WKB scenario here discussed, it is possible that a pre-inflationary information survives to the de Sitter expansion of the Universe. In fact, the inhomogeneous component of the tensor perturbation is not fixed by the slow-rolling evolution, but it contains, in the Fourier transform of the original space dependence, an arbitrary shape, in principle corresponding to non-gaussian features of the tensor spectrum, to be identified by future experiments aimed to measure the cosmological background of gravitational waves \cite{Campeti_2021} (for a review of the main mechanisms of gravitational wave production related to inflation see \cite{GWBdetection2}). 

\vspace{2mm}

The manuscript is structured as follows. In Sec. \ref{sec:QIsolution} we present the quasi-isotropic solution, both in the presence of ultrarelativistic matter and during the de Sitter phase. In Sec. \ref{sec: Vilenkin} we review the quantization scheme à la Vilenkin. In Sec. \ref{sec:GeneralDynamics} we construct the gauge invariant action of the quasi-isotropic Universe in presence of a real self-interacting scalar field. We discuss the different nature of the scalar and tensor degrees of freedom and derive the total Hamiltonian of the background and perturbations. Then, in Sec. \ref{sec:Quantumdynamics} we quantize the perturbations implementing the semiclassical and Born-Oppenheimer-like approximation. We derive a time dependent Schr\"odinger equation for the wave function of each independent mode and find the solution. Finally, we derive the Power Spectrum of both the scalar and tensor perturbations first in a purely de Sitter expansion and then in a generic slow-roll inflationary scenario. In Sec. \ref{sec:Conclusions} some concluding remarks and future perspectives follow.

\section{\label{sec:QIsolution}Quasi-isotropic solution}


The original Lifshitz-Khalatnikov quasi-isotropic solution \cite{LKoriginal,lifshitz1963investigations} considers an inhomogeneous Universe filled by ultrarelativistic matter nearby the cosmological singularity. The peculiar feature of this solution consists in the fact that the inhomogeneity is managed through a Taylor expansion of the spatial metric $\gamma_{ij}(t,\textbf{x})$ in the synchronous time\footnote{In the following Latin indices run through the three spatial values 1, 2, and 3, while Greek indices run through the values 0, 1, 2 and 3.}. The idea is thus that, order by order, space contracts (going backward in time) maintaining the same time dependence of linear distance changes.

Because the FLRW solution for a relativistic perfect fluid with equation of state $p=\rho/3$ is linear in the synchronous time $t$, a quasi-isotropic extension of this geometry should consists of an expansion of the spatial metric in integer powers of $t$. Therefore, limiting ourselves to the deal only with the first two terms of expansion and neglect higher order contributions, we consider the following metric
\begin{equation}\label{LKMetricExpansion}
    \gamma_{ij}(t,\textbf{x}) = h_{ij}(\textbf{x})t + \theta_{ij}(\textbf{x})t^2 .
\end{equation}
The functions $h_{ij}$ are chosen arbitrary while the functions $\theta_{ij}$ - as well as the energy and velocity distributions for matter - can be expressed through these functions. Moreover, all the operations of index raising and covariant differentiation are made using the time-independent tensor $h_{ij}$, i.e. it takes a metric role. For example $\theta^i_j=h^{ik}\theta_{kj}$. To lowest order, the inverse matrix of the metric and the determinant read as
\begin{equation}\label{LKmetric}
    \gamma^{ij} = \frac{h^{ij}}{t}  - \theta^{ij}, \qquad \gamma = j\,t^3\Big(1+\frac{1}{2}\theta t  \Big),
\end{equation}
where $\gamma = \det(\gamma_{ij})$ and $j=\det(h_{ij})$. 

In the presence of an ultrarelativistic perfect fluid with energy density $\rho_r$ and four-velocity $u^\mu$ the Einstein field equations, in the synchronous frame, assume the form
\begin{subequations}\label{eq:GravEquations}
\begin{align}
    &\frac{1}{2}\partial_tk^i_i - \frac{1}{4}k^i_jk^j_i =\kappa_g\frac{\rho_r}{3} (4u_0^2+1), \label{Eeq00} \\[0.2cm]
    &\frac{1}{2}(\nabla_ik^j_{j}-\nabla_jk^j_{i})=\frac{4}{3}\kappa_g\rho_r u_iu^0, \label{Eeq0i}\\[0.2cm]
    &R^i_j + \frac{1}{2\sqrt{\gamma}}\partial_t(\sqrt{\gamma}k^i_j) = \kappa_g\frac{\rho_r}{3} (4u_ju^i+\delta^i_j), \label{Eeqij}
\end{align}
\end{subequations}
where $\kappa_g = 8\pi G$ denotes the Einstein constant (we set $c=1$), $k_{ij} = \partial_t \gamma_{ij}$ and $R^i_j = h^{ik}\,{R_{kj}}$ represents the three-dimensional Ricci tensor obtained by the time-independent metric $h_{ij}$. 

The solutions of the system \eqref{eq:GravEquations} are constructed asymptotically in the limit $t\rightarrow0$. From Eqs. \eqref{Eeq00} and \eqref{Eeq0i} and considering the identity $-1 = u_\mu u^\mu \sim -u^2_0 + t^{-1}h^{ij}u_iu_j$, one finds that the first two terms of the energy density expansion and the leading term of the velocity are given by
\begin{align}
    &\kappa_g\rho_r = \frac{3}{4t^2} - \frac{\theta}{2t}, \label{QI3} \\[0.2cm]
    &u_i = \frac{t^2}{2}(\nabla_i\theta-\nabla_j\theta^j_i). \label{QI4}
\end{align}
As a consequence, the density contrast $\delta$ can be expressed as the ratio between the first and zeroth-order energy density terms, i.e. $\delta \sim\theta t$. Therefore, in agreement with the standard cosmological model, the homogeneous leading term of energy density diverges more rapidly than the perturbations and the singularity is naturally approached with a vanishing density contrast in this scenario.

The leading order of the three-dimensional Ricci tensor can be written as $R^i_j = \tilde R^i_j/t$, where $\tilde R_{ij} = h_{ik}\tilde R^k_j$ denotes the Ricci tensor corresponding to $h_{ij}$. This causes the terms of order $t^{-2}$ in eq. \eqref{Eeqij} identically cancel out, while those proportional to $t^{-1}$ give the following relation between the six arbitrary functions $h_{ij}$ and the coefficients $\theta_{ij}$
\begin{equation}
    \theta^i_j = -\frac{4}{3}\tilde R^i_j + \frac{5}{18}\tilde R\delta^i_j.
\end{equation}

Now, using the three-dimensional Bianchi identity $\nabla_i\tilde R^i_j = \frac{1}{2}\nabla_j \tilde R$, one obtains that the spatial dependence of $u_i$ in Eq. \eqref{QI4} reduces to the divergence of $\theta$. This means that (at this level of approximation) the curl of the velocity vanishes and no rotations take place in the fluid.

Finally, it must be observed that the metric \eqref{LKMetricExpansion} admits an arbitrary transformation of the three-space coordinates, therefore the number of arbitrary space functions arising from $h_{ij}$ is $6 - 3 = 3$. The pure isotropic and homogeneous model results to be a special case of this solution, for $h_{ij}$ corresponding to spaces of constant curvature, i.e. $\tilde R^i_j = \text{const} \times \delta^i_j$.

\subsection{Inflationary scenario}\label{subsec:QIsolution-inflation}

The scheme previously outlined can be applied also in the context of inflation, with two differences: (i) the presence of a scalar field allows to relax the assumption of expandability in integer powers adopted in eq. \eqref{LKMetricExpansion} and (ii) the asymptotic solution is searched in the limit $t \rightarrow \infty$. Therefore, we require a three-dimensional metric tensor having the following structure
\begin{equation}
\begin{split}\label{qi space metric}
    \gamma_{ij}(t,\textbf{x}) = a^2(t) [h_{ij}(\textbf{x}) + \eta(t)\theta_{ij}(\textbf{x}) ],
\end{split}
\end{equation}
where $\eta = b^2/a^2$ and is assumed to satisfy the condition
\begin{equation}
    \lim_{t\rightarrow\infty}\eta(t) = 0.
\end{equation}
As shown in Ref. \cite{montani2003quasi}, the coupling between the metric \eqref{qi space metric}, a scalar field $\phi(t,\textbf{x})$ governed by an effective cosmological constant $\Lambda$ and an ultrarelativistic perfect fluid yields an isotropic and exponentially growing flat background metric
\begin{equation}\label{MontaniBG}
    a(t) = a_0 e^{H_0t}, \qquad h_{ij}=\delta_{ij},
\end{equation}
perturbed by an exponentially decreasing inhomogeneous term. i.e. 
\begin{equation}\label{MontaniPerturbation}
    \eta(t) = \eta_0 e^{-4H_0t}.
\end{equation}
In Eqs. \ref{MontaniBG} and \eqref{MontaniPerturbation} $H_0=\sqrt{\kappa_g\Lambda/3}$ is the (constant) Hubble parameter. The energy density and velocity of the ultrarelativistic field are given by
\begin{equation}\label{MontaniMatter}
    \rho_{r} = -\frac{4}{3}\Lambda\eta\theta, \qquad u_i= -\frac{1}{4H_0}\partial_i|\log\theta|,
\end{equation}
which implies $\theta < 0$ for each point of the allowed domain of the spatial coordinates. Furthermore, the spatial distribution of the metric correction reduces to $\theta_{ij} = \theta\delta_{ij}/3$. Finally, the two leading orders of the scalar field, modulo a multiplicative integration constant, read as
\begin{equation}\label{MontaniScalarField}
    \phi \propto \sqrt{\frac{t_r}{t_r-t}}\Big( 1 - \frac{1}{12H_0}\frac{\eta}{t_r-t}\theta \Big),
\end{equation}
where $t_r$ is a constant which satisfies the condition $t_r\gg t$ during all the slow-rolling expansion. This condition guarantees that the dominant term of the scalar field is almost constant such that it actually evolves very slowly on the plateau region.

The peculiar feature of this solution lies in the free character of the function $\theta(\textbf{x})$ which, being a three-scalar, is not affected by spatial coordinate transformations. From a cosmological point of view, this implies the existence of a quasi-isotropic inflationary solution together with an arbitrary spatial distribution of ultrarelativistic matter and of the scalar field. Nevertheless, given the extremely fast decaying of the inhomogeneities, they are not able to account for the observed cosmological structures\footnote{Given that the minimum number of e-folds necessary to solve the shortcomings present in the standard cosmological model is typically $\mathcal{E} \simeq 60$ \cite{kolb2018early}, any perturbation that could exist before inflation would be reduced at the end by a factor $\sim \eta_f/\eta_i \sim (a_i/a_f)^4 \sim \mathcal{O}(10^{-108})$.}, therefore supporting the idea that the the latter cannot have a classical origin in the presence of an inflationary scenario.

\section{Semiclassical approximation} \label{sec: Vilenkin}

During the slow-roll phase the Universe is essentially classical and the quantum fluctuations of the inflaton and gravitational fields evolve as small quantum fields around a classical FLRW background \cite{mukhanov1992theory}. This situation is particularly suitable to implement the scheme proposed by Vilenkin in \cite{vilenkin1989interpretation}. The idea is that in cosmology a well-defined evolutionary quantum dynamics can emerge from the timeless WDW equation only when some of the degrees of freedom of the Universe become classical. 
The reason for that is due to the essential role played by classical measuring devices in quantum mechanics. The existence of a classical background during inflation would thus ensure the presence of an “observer” of the quantum subsystem represented by the fields fluctuations.

\vspace{5mm}

Here we restrict the discussion to homogeneous minisuperspace models, but we will see that the formalism can be straightforwardly apply to our model which includes also a certain degree of inhomogeneity. In this case, the geometric and matter variables are independent of $\textbf{x}$ and the momentum constraint identically vanishes. The quantum dynamics of the Universe follows from the WDW equation \cite{wheeler1968battelle,dewitt1967quantum}
\begin{equation}\label{Vilenkin-WDW}
    [\hbar^2\nabla^2 - U(h)]\Psi(h) = 0,
\end{equation}
where we used a unified notation for metric and matter variables, all labeled as $h^\alpha$, and $\nabla^2$ denotes the Laplace-Beltrami operator of the superspace metric $G_{\alpha\beta}$. The physical states are the solutions of the WDW equation \eqref{Vilenkin-WDW} and $\Psi = \Psi(h)$ is called the \textit{wave function of the Universe}.

However, in a fully quantum regime two main problems arise \cite{isham1993canonical}: it is not clear how to recover an explicit evolution with respect to a time variable (the so called \textit{problem of time}); in general a satisfactory definition of probability associated to $\Psi(h)$ has not yet been found and, therefore, it is not clear what is its meaning. Indeed, since eq. \eqref{Vilenkin-WDW} is formally a Klein-Gordon equation with a variable mass, it is common to consider the corresponding Klein-Gordon current
\begin{align}\label{KGcurrent}
    j^\alpha = -i\frac{\hbar}{2}G^{\alpha\beta}(\Psi^*\nabla_\beta\Psi-\Psi\nabla_\beta\Psi^*), \quad \nabla_\alpha j^\alpha = 0.
\end{align}
This current gives the probability $dP=j^\alpha d\Sigma_\alpha$ to find the system in a surface element $d\Sigma_\alpha$ of the configuration-space. The problem is that, in this context, this quantity is not positive-defined in general. 

Both problems can be solved if we assume the system to be composed of $n-m$ semiclassical variables and $m$ quantum variables, that we label respectively $\lbrace c^\alpha \rbrace$ and $\lbrace q^\nu \rbrace$. Furthermore, we suppose that the quantum subsystem can be considered “small" (so that its backreaction on the classical variables is negligible) and “fast" respect to the classical one. In this way it is possible to implement a Born–Oppenheimer approximation to decouple the quantum and classical dynamics. We then write the WDW equation \eqref{Vilenkin-WDW} in the following convenient form
\begin{align}\label{Vilenkin-WDW2}
    \big( \hbar^2\nabla_0^2 - U_0(c) - H_q \big)\Psi = 0, 
\end{align}
where $H_{c} = -\hbar^2\nabla_0^2 + U_0$ is obtained neglecting all the contributes of the quantum variables. We can also write the wave function as
\begin{equation}
    \Psi = A(c)e^{iS(c)/\hbar}\chi(c,q).
\end{equation}

As shown in \cite{vilenkin1989interpretation}, the two leading orders in $\hbar$ of the WKB expansion of Eq. \eqref{Vilenkin-WDW2}, considering also that the wave function $\psi_0 = Ae^{iS/\hbar}$ satisfies the equation $H_c\psi_0=0$, lead to: 

\vspace{5mm}

\noindent 1) The Hamilton-Jacobi equation
    \begin{equation}
        G^{\alpha\beta} \nabla_\alpha S \, \nabla_\beta S + U_0 = 0.
    \end{equation}
$S=S(c)$ is the classical action and describes a congruence of classical trajectories. It is thus possible to define a time variable $t$ which parametrizes the evolution of each trajectory. 

\vspace{5mm}

\noindent 2) The Schr\"{o}dinger equation for the quantum subsystem
\begin{equation}
    i\hbar\frac{d\chi}{dt} = NH_q\chi,
\end{equation}
where now the quantum wave function $\chi = \chi(t,q)$ depends on time $t$.

\vspace{5mm}

\noindent 3) An equation for the amplitude $A$ expressing the conservation of the classical probability current $j_0^\alpha = |A|^2\nabla_0^\alpha S$ for the variables $h^\alpha$, obtained from the leading order of Eq. \eqref{KGcurrent}. In this way one gets two different parts of the probability current, one related to
the components of the classical subspace and the other
one to those in the quantum one, ending with a normalizable probability distribution for the quantum variables $q^\nu$, given by $\rho_\chi= |\chi|^2$.

\vspace{5mm}

Hence, the standard interpretation of the wave function has been recovered for a small subsystem of the Universe.

\section{\label{sec:GeneralDynamics}Action functional}
 
Now lets consider a quasi-isotropic scenario in presence of a real and self-interacting scalar field $\Phi$. The Einstein-Hilbert action for the gravity and matter system under consideration is\footnote{We consider the signature of the metric to be $(-,+,+,+)$.}
\begin{equation}\label{General action}
    S = \int d^4x \sqrt{-g} \Big[\frac{1}{2\kappa_g}R - \frac{1}{2}g^{\mu\nu}\nabla_\mu\Phi\nabla_\nu\Phi - V(\Phi)\Big].
\end{equation} 

We assume that the contribution of the inhomogeneities is small enough to treat them as a first order correction to the FLRW background, so that the metric and scalar fields can be written as 
\begin{align}
 &g_{\mu\nu}(t,\textbf{x}) = g^{RW}_{\mu\nu}(t) + \delta g_{\mu\nu}(t,\textbf{x}),\label{metric1}\\[0.3cm]
    &\Phi(t,\textbf{x}) = \phi(t) + \varphi(t,\textbf{x}).
\end{align}
The metric of the spatial slicing  $\gamma_{ij}$ takes the form \eqref{qi space metric}, while now the other components can not be fixed, as in the case seen in section \ref{sec:QIsolution}, because they will be subsequently quantized. In addition, since we are interested in the inflationary dynamics, we assume from the beginning a background metric with null curvature.

Expanding the action \eqref{General action} in $\delta g_{\mu\nu}$ and $\varphi$, the background dynamics and the perturbations dynamics follow respectively from the zeroth and second order terms of the expansion, since they give the unperturbed and linearized Einstein equations \cite{taub1969stability,mukhanov1992theory}. Therefore, the steps toward the quantization of the perturbations are: expand the action \eqref{General action} up to second order in perturbations, find the super-Hamiltonian constraint and apply Vilenkin's procedure discussed in Sec. \ref{sec: Vilenkin}.

\subsection{\label{subsec:GIvariables}Gauge invariant variables}

In order to obtain the canonical action we need to write \eqref{General action} in terms of the ADM variables \cite{cianfrani2014canonical}. Comparing the metric \eqref{metric1} with the line element
\begin{equation}
    ds^2 = -\big(\mathcal{N}^2 - \mathcal{N}_i\mathcal{N}^i\big)dt^2 + 2\mathcal{N}_idx^idt + \gamma_{ij}dx^idx^j,
\end{equation}
we obtain the lapse function $\mathcal{N}$ and shift vector $\mathcal{N}^i$ (note that each quantity must be expand up to second order in the perturbation variables)
\begin{align}
    &\mathcal{N} = N + \delta N - \frac{\delta N^2-\gamma_{ij}\delta N^i\delta N^j}{2N},\\[0.2cm]
    &\mathcal{N}_i = \delta N_i,
\end{align}
where we have defined the variables
\begin{equation}
    N=\sqrt{-g^{RW}_{00}}, \ \  \delta N = -\frac{\delta g_{00}}{2\sqrt{-g^{RW}_{00}}}, \ \  \delta N_i = \delta g_{0i}.
\end{equation}

From the condition $\gamma^{ik}\gamma_{kj}=\delta^i_j+\mathcal{O}(\eta^3)$ we get the inverse three-metric tensor\footnote{From now on we will adopt the convention $\delta^{ij}V_iW_j = V_iW_i$, while $\gamma^{ij}V_iW_j = V^iW_i$.}
\begin{equation}
    \gamma^{ij} = \frac{1}{a^2} \big( \delta_{ij} - \eta\theta_{ij} + \eta^2\theta_{ik}\theta_{kj} \big).
\end{equation}
Finally, the squared root of the spatial metric’s determinant reads as 
\begin{equation}
    \sqrt{\gamma} = a^3 \Big[ 1 + \frac{1}{2}\eta\theta + \frac{1}{8}\big(\theta^2-2\theta_{ij}\theta_{ij}\big) \Big].
\end{equation}

Only the scalar and tensor components of the perturbations ``survive" the inflationary expansion, so that we can write
\begin{align}
    &\theta_{ij} = -2\psi\delta_{ij} + 2\mu_{,ij} + \vartheta_{ij}\\[0.3cm]
    &\delta N_i = aB_{,i},
\end{align}
where $\psi(\textbf{x}),\ \mu(\textbf{x}),\ B(t,\textbf{x})$ are scalar functions, $(...)_{,i}=\partial_i(...)$ and $\vartheta_{ij}(\textbf{x})$ is a symmetric, traceless and divergenceless three-tensor.

While the tensor sector of the perturbations (which represents the gravitational degrees of freedom $\delta g_{ij}^{(t)} = \eta\vartheta_{ij}$) is gauge invariant, 
the scalar component is not\footnote{We recall that in the context of cosmological perturbation theory this term denotes those coordinate transformations which involve only perturbative quantities: $x^\mu\rightarrow x^\mu+\epsilon^\mu$, where $\epsilon$ is of the order of $\delta g_{\mu\nu}$ and $\varphi$.}. However, it is much more convenient to work with only gauge invariant variables, which represent the true physical degrees of freedom among the perturbations. These variables will in fact trivialize the generators of the gauge transformations, i.e. the first-order perturbation of the super-Hamiltonian and supermomentum constraints, thus removing the gauge artifacts. 

The gauge invariant degree of freedom of the scalar sector can be constructed combining the two Bardeen's potentials \cite{Bardeen1980} with the scalar field perturbation, obtaining the variable
\begin{equation}\label{ScalarGIvariable}
    \xi_S(t,\textbf{x}) = \varphi(t,\textbf{x}) + \frac{\dot\phi}{H}\eta(t)\psi(\textbf{x}).
\end{equation}
However, in practice it is more useful to work with an auxiliary field defined as
\begin{equation}\label{MukhanovVariableS}
    v_S(t,\textbf{x}) = a\Big[\varphi(t,\textbf{x}) + \frac{\dot\phi}{H}\eta(t)\psi(\textbf{x})\Big],
\end{equation}
which is called the Mukhanov-Sasaki variable \cite{mukhanov1992theory}.

Let us point out that $\xi_S$ and $v_S$ factorize into a product of a space and a time function only if $\varphi$ also does. Nevertheless, this does not happen in general as it is easy to see looking at the de Sitter limit $\dot\phi \rightarrow 0$. In this case $\varphi$ does not couple to the metric (the gauge invariant variables \eqref{ScalarGIvariable} and \eqref{MukhanovVariableS} become independent from it) and there is no reason to impose any factorization. We will see that at a quantum level there is no chance to recover the quasi-isotropic factorization for the scalar sector. Differently the tensor component is automatically gauge invariant because it does not appear in the perturbative constraints. This means, in particular, that the quasi-isotropic factorization is gauge invariant so that the tensor inhomogeneities evolve with a frozen spatial distribution $\vartheta_{ij}(\textbf{x})$. 

This difference is crucial because, as shown in Sec. \ref{sec:Quantumdynamics}, it implies that the quantum behavior of the scalar and tensor degrees of freedom is completely different.

\subsection{Gauge invariant action}\label{subsec:GIaction}

\subsubsection{Lagrangian formulation}

The expansion of the action \eqref{General action} in the perturbations $\delta g_{\mu\nu}$ and $\varphi$,
\begin{equation}
    S=S_0+\delta S+\delta^2S+...,
\end{equation} 
gives the action $S_0$ of the background variables $a$ and $\phi$, its first-order variation $\delta S$ and the action $\delta^2S$ of the perturbations. The latter splits into the sum of two terms, one containing only scalar degrees of freedom, $S_S$, and one only the tensor ones, $S_T$.
	
Using the conformal time $d\tau=dt/a$ and the notation $'=d/d\tau$, $S_0$ results to be the usual FLRW action in presence of a scalar field 
\begin{equation} \label{SBG}
	S_0 = \int d\tau\,\Big[ -\frac{3a'^2}{\kappa_g N} + \frac{a^2\phi'^2}{2N} - Na^4V(\phi) \Big].
\end{equation}
The fiducial length $l=\big(\int d^3x\big)^{1/3}$ has been absorbed by the following redefinition of the scale factor and conformal time\footnote{Note that because the interval $ds$ has dimension of a length, we can initially choose $[a]=L^0$, $[dx]=[d\tau]=L$ and thus $[l] = L$, and so after the rescaling \eqref{scalefactorRedefinition} the scale factor has the dimension of a length and $\tau$ and the spatial variables become dimensionless.}
\begin{equation}\label{scalefactorRedefinition}
	a \rightarrow la, \qquad \tau \rightarrow \tau/l.
\end{equation}
In fact, $l$ has to be included in the action since we are considering both homogeneous and inhomogeneous quantities. 

Applying the background equation of motion derived from \eqref{SBG} to the unperturbed variables $a$ and $\phi$ makes, as expected, the first order term $\delta S$ identically vanish. Furthermore, it simplifies a lot the second-order term $\delta^2S$. 

The scalar component $S_S$ contains the first variation of the background constraints. Imposing the constraint given by varying $S_S$ with respect to the metric variable $B$ and using the gauge invariant variable \eqref{MukhanovVariableS} the scalar perturbation action results to have the following expression 
\begin{widetext}
\begin{equation}\label{Saction}
    S_S = \int d\tau d^3x \Big[ \frac{1}{2N}v_S'^2 - \frac{N}{2}v_{S,i}v_{S,i} + \frac{N}{2}\Big( \frac{2\mathscr{H}^2}{N^2} - \frac{7\kappa_g{\phi'}^2}{2N^2}  + \frac{\kappa_g^2{\phi'}^4}{2\mathscr{H}^2N^2} - \frac{2a^2\kappa_g{\phi'}}{\mathscr{H}}V_{,\phi} - a^2V_{,\phi\phi} \Big)v_S^2 \Big],
\end{equation}
\end{widetext}
where $\mathscr{H}=a'/a$. It is thus useful to write Eq. \eqref{Saction} in Fourier space. Indeed, the quantization of the perturbations becomes simpler if one considers the Fourier amplitudes 
\begin{equation}
	v_\textbf{k}(\tau) = \int \frac{d^3x}{(2\pi)^{3/2}} v_S(\tau,\textbf{x}) e^{-i\textbf{k}\cdot\textbf{x}}.
\end{equation}
Here, $\textbf{k}$ is the comoving wave-vector, related to the physical wave-vector by $\textbf{k}_{phys} = \textbf{k}/a$, and the condition $v_{-\textbf{k}}=v_{\textbf{k}}^*$ must hold because $v_S(\tau,\textbf{x})$ is real. The complicated functional Schr\"odinger equation for the wave functional $\psi[\tau,v(x)]$ actually can be reduced to an infinite number of independent Schr\"odinger equations for the single modes. 

To achieve this result we need to replace the integral over the wave-vector $\textbf{k}$ by a sum, i.e. $\int d^3k \ \rightarrow \ \sum_\textbf{k}/l^3$. In Sec. \ref{sec:Quantumdynamics} we will see that the equations of motion and the initial conditions for the modes will only depend on the module $k$. Therefore, the sum over the wave-vector has to be taken over the module $k$. In this way the action \eqref{Saction} becomes 
\begin{equation}\label{GISAction}
	S_S = \int d\tau\, \frac{1}{l^3}\sum_{k}\Big[ \frac{1}{N}v_{k}'{v^*_{k}}' - N\omega_{k}^2 v_{k}v^*_{k} \Big],
\end{equation}
where we have defined the quantities
\begin{equation} \label{scalarfreq}
\begin{split}
	\omega_{k}^2 = k^2 - \frac{2\mathscr{H}^2}{N^2} + \frac{7\kappa_g{\phi'}^2}{2N^2} - \frac{\kappa_g^2{\phi'}^4}{2\mathscr{H}^2N^2} \\
	+ \frac{2a^2\kappa_g{\phi'}}{\mathscr{H}}V_{,\phi} + a^2V_{,\phi\phi},
\end{split}
\end{equation}
which are functions of the unperturbed variables only. We will see later that they take the role of time dependent frequencies.  
	
The tensor perturbation action obtained from the expansion of Eq. \eqref{General action} results to be in a gauge invariant form. Thanks to the quasi-isotropic factorization the spatial dependence of the Lagrangian density can be integrated in the action. All the information about the spatial distribution of the inhomogeneities results to be encoded into the two constants $\mathcal C_\vartheta = \int d^3x \vartheta_{ij}\vartheta_{ij}$ and $\mathcal C_\vartheta' = \int d^3x \vartheta_{ij,k}\vartheta_{ij,k}$. Therefore, we are left with just one degree of freedom in the action. In analogy to Eqs. \eqref{ScalarGIvariable} and \eqref{MukhanovVariableS}, we now define the tensor gauge invariant variables
\begin{equation}\label{MukhanovVariableT}
    \xi_T = \sqrt{\frac{\mathcal C_\vartheta}{4\kappa_g}}\eta, \qquad v_T =  \sqrt{\frac{\mathcal C_\vartheta}{4\kappa_g}}a\eta. 
\end{equation}
In this way the tensor perturbation action becomes the following
\begin{equation}\label{GITAction}
	S_T = \int d\tau \Big[ \frac{1}{2N}v_T'^2 - \frac{N}{2}\omega_T^2v_T^2 \Big].
\end{equation}
In this case the time dependent frequency of the tensor degree of freedom is given by
\begin{equation}\label{tensorfreq}
    \omega_T^2 = K^2 - \frac{2\mathscr{H}^2}{N^2} + \frac{\kappa_g {\phi'}^2}{2N^2}, 
\end{equation}
where $K =\sqrt{\mathcal C_\vartheta'/\mathcal C_\vartheta}$. The meaning of the constant $K$ results clearer in Fourier space: it is the square root of the expectation value of $k^2$ with respect to the probability distribution\footnote{It is easy to see that $\mathcal P_\vartheta$ satisfies all the requirement to be a probability distribution, in particular the three Kolmogorov axioms \cite{kolmogorov}.} $p_\vartheta(\textbf{k})=|\hat\vartheta_{ij}(\textbf{k})|^2/\int d^3k|\hat\vartheta_{ij}|^2$. Indeed, it results to be equal to
\begin{equation}
	K = \Big[ \int d^3k p_\vartheta(\textbf{k}) k^2 \Big]^{1/2}= \braket{k^2}_{\vartheta}^{1/2}.
\end{equation}
We shall see that this quantity represents the reference scale for the evolution of the tensor degree of freedom \eqref{MukhanovVariableT}, playing a role equivalent of the wave vector $k$ for the scalar mode-$k$.
	
As for the background action we remove the explicit appearance of the fiducial length $l$ in the perturbations action by the replacement \eqref{scalefactorRedefinition} and the following redefinitions
\begin{subequations}
\begin{align}\label{variablesRedefinition}
	&v_{k} \rightarrow \frac{v_{k}}{l^2}, \qquad v_T \rightarrow \frac{v_T}{\sqrt{l}}, \\[0.2cm]
	&{k} \rightarrow l{k}, \qquad K \rightarrow lK,
\end{align}
\end{subequations}
so that ${k}$ and $K$ become dimensionless.
	
Finally, the total action governing both unperturbed and perturbed dynamics is given by the sum of Eqs. \eqref{SBG}, \eqref{GISAction} and \eqref{GITAction}.

\subsubsection{Hamiltonian formulation}

In order to quantize the system, it is convenient to replace the complex amplitudes $v_{k}$ with real variables given by their real and imaginary part 
\begin{equation}\label{RealImaginaryparts}
    v^R_{k}= \sqrt{2}\Re (v_{k}), \qquad v^I_{k}= \sqrt{2}\Im (v_{k}).
\end{equation}
Defining from the Lagrangians \eqref{SBG}, \eqref{GISAction} and \eqref{GITAction} the canonical momenta 
\begin{equation}\label{momenta}
\begin{split}
	&\pi_a =  -\frac{6{a'}}{\kappa_g N}, \qquad\ \  \pi_\phi = \frac{a^2{\phi'}}{N}, \\[0.2cm]
	&\pi^i_{{k}} = \frac{{v^i_{k}}'}{N}, \qquad\ \ \ \ \ \  \pi_T = \frac{v_T'}{N} ,
\end{split}
\end{equation}
where the index $i$ labels the real and imaginary components \eqref{RealImaginaryparts} of the scalar modes, we obtain an Hamiltonian $H=N\mathcal{H} = N(\mathcal{H}_0+\mathcal{H}_S+\mathcal{H}_T)$ which contains the background constraint $\mathcal{H}_0$ and its second variation $\delta^2\mathcal{H} = \mathcal{H}_S+\mathcal{H}_T$, given in the following expressions 
\begin{subequations}
\begin{align}
    &\mathcal{H}_0 = -\frac{\kappa_g}{12}\pi_a^2 + \frac{1}{2a^2}\pi_\phi^2 + a^4V(\phi), \\[0.2cm]
    &\mathcal{H}_S = \sum_{i=R,I}\sum_{{k}}\Big(\frac{{\pi^i_{{k}}}^2}{2} + \frac{\omega_{k}^2}{2}{v^i_{{k}}}^2\Big),\label{Sconstraint}\\[0.2cm]
    &\mathcal{H}_T =  \frac{\pi_T^2}{2} + \frac{{\omega_T^2}}{2}v_T^2\,.\label{Tconstraint}
\end{align} 
\end{subequations}

Let us observe that once the frequencies \eqref{scalarfreq} and \eqref{tensorfreq} are expressed in terms of the momenta \eqref{momenta}, the unperturbed component $N$ of the lapse function its removed from the super-Hamiltonian $\mathcal{H}$, thus appearing in the Hamiltonian only as a multiplicative factor. Therefore, we obtain the super-Hamiltonian constraint 
\begin{equation}\label{super-Hamiltonian constraint}
	\mathcal{H}=\mathcal{H}_0+\mathcal{H}_S+\mathcal{H}_T=0.
\end{equation} 

Note that here we are considering a total Hamiltonian for both the background and perturbative degrees of freedom. In this way, what we are going to do, is to directly quantize all degrees of freedom with the total constraint \eqref{super-Hamiltonian constraint}. This is, in principle, different from solving separately the dynamics of the background and then the dynamics of the perturbations. In this second case there would be
 two different Hamiltonians, the first containing only unperturbed variables and the other terms quadratic in the perturbations. The background variable $N$ would appear in the Hamiltonian of the perturbations as a given, non-dynamical variable, so that $H_{pert}=N(\mathcal{H}_S+\mathcal{H}_T)$ would represent a physical (i.e. non-vanishing) Hamiltonian for the perturbative degrees of freedom. Therefore, once quantized, it would lead to an evolutive Schr\"odinger equation for the perturbations. 
 
 Nonetheless, as long as we are interested in the linear regime for the perturbations and in the limit of quantum field theories on a fixed background spacetime, the two approaches are equivalent. However, we stress that only the method used here is able to give the quantum-gravitational corrections to the background dynamics \cite{Kiefer1,Kiefer2}.

\section{\label{sec:Quantumdynamics}Quantum dynamics}

\subsection{\label{subsec:Quantumdynamics}Wave function of the perturbations}

Let us now face the problem of quantization. The phase space variables are promoted to quantum operators which satisfy the canonical commutation relations 
\begin{equation}
\begin{split}
	&[a,\pi_a] = i, \qquad \ \ \ \ \ \ \ \ \ \, [\phi,\pi_\phi] = i, \\ 
	&[v^i_{{k}},\pi^j_{{k'}}] = i\delta_{ij}\delta_{{k},{k'}}, \quad \ [v_T,\pi_T] = i.
\end{split}
\end{equation}
In the following it will be convenient to introduce a unified notation for the scalar and tensor modes, labelling all the perturbations as $q_\nu$. 

The super-Hamiltonian constraint \eqref{super-Hamiltonian constraint} now becomes a quantum operator which annihilates the wave functional $\psi=\psi(a,\phi,\{q_\nu\})$, yielding the WDW equation $\mathcal{H}\psi=0$.

Now we apply the semiclassical scheme à la Vilenkin. We treat the background degrees of freedom $a$ and $\phi$ as classical and impose the following WKB ansantz
\begin{equation}
    \psi = A(a,\phi)e^{iS(a,\phi)/\hbar}\chi(a,\phi,\{q_\nu\}).
\end{equation}
The wave function $\psi_0 = Ae^{iS/\hbar}$ satisfies the equation $\mathcal{H}_0\psi_0=0$
which gives, at the zeroth-order in $\hbar$, the Hamilton-Jacobi equation
\begin{equation}\label{H-Jeq}
	-\frac{\kappa_g}{12} \Big(\frac{\partial S}{\partial a}\Big)^2 + \frac{1}{2a^2} \Big(\frac{\partial S}{\partial\phi}\Big)^2 + a^4V = 0.
\end{equation}
One can demonstrate that this equation is equivalent to the Friedmann equation. Consequently, the variables $a$ and $\phi$ follow classical trajectories with ``velocities" given by
\begin{equation}\label{velocities}
\begin{split}
   & a' = N\frac{\partial \mathcal{H}_0}{\partial \pi_a} = -\frac{\kappa_g N}{6}\frac{\partial S}{\partial a}\\[0.3cm]
   & \phi' = N\frac{\partial \mathcal{H}_0}{\partial \pi_\phi} = \frac{N}{a^2}\frac{\partial S}{\partial \phi}.
\end{split}
\end{equation}
From equations \eqref{velocities} it is then possible to define a WKB time in terms of the background variables. This time variable is given by the following relation
\begin{equation}\label{WKBtime}
    \frac{\partial}{\partial \tau} = N\Big(-\frac{\kappa_g}{6}\frac{\partial S}{\partial a}\frac{\partial}{\partial a} + \frac{1}{a^2} \frac{\partial S}{\partial \phi}\frac{\partial}{\partial \phi}\Big),
\end{equation}
and we shall show that it will be identified with the classical conformal time. 

Applying the definition \eqref{WKBtime} and considering the Born-Oppenheimer approximation discussed in Sec. \ref{sec: Vilenkin}, we find that the leading order in $\hbar$ of the WDW equation gives the following Schr\"odinger equation for the quantum subsystem
\begin{equation}\label{Vilenkin3}
    i\hbar \frac{\partial\chi}{\partial \tau} =  NH_q\chi,
\end{equation}
where now $\chi=\chi(\tau,\{q_\nu\})$ and $H_q$ is the Hamiltonian of the perturbations
\begin{equation}
    H_q = \sum_\nu \Big(-\frac{\hbar^2}{2}\frac{\partial^2}{\partial q_\nu^2} + \frac{\omega_\nu^2}{2}q_\nu^2 \Big).
\end{equation}
Let us note that the semiclassical approximation simplifies drastically the expression of the frequencies $\omega_{k}$ and $\omega_T$. In a fully quantum gravitational regime, because of the presence of the unperturbed momenta, they are quantum operators. However, at the order of approximation considered here, the results do not change if we still use their classical expressions.

It must be observed that, in order to solve equation \eqref{Vilenkin3}, we need to fix the lapse function and solve the dynamics of the background so that the frequencies $\omega_k$ and $\omega_T$ can be expressed as functions of time. This means that we must fix the reference frame, which is now possible thanks to the semiclassical regime. We thus choose to work in a synchronous frame and set $N(t) = 1$.

Thanks to the linear character of the perturbations it is possible \cite{kiefer1987continuous} to simplify equation \eqref{Vilenkin3}. In this regime different modes do not interact with each other and they can be considered independent. This assumption allows us to factorize $\chi$ into a product of wave functions describing the single modes, that is
\begin{equation}
    \chi(\tau,\{q_\nu\}) = \prod_\nu\chi_\nu(\tau,q_\nu).
\label{WFansatz}
\end{equation}
In this way equation \eqref{Vilenkin3} reduces to an infinite set of decoupled Schr\"odinger equations which read as
\begin{equation}\label{Schroedinger}
    i\hbar \frac{\partial\chi_\nu}{\partial \tau} =  \Big(- \frac{\hbar^2}{2}\frac{\partial^2}{\partial q_\nu^2}+ \frac{\omega_\nu^2}{2}q_\nu^2\Big)\chi_\nu.
\end{equation}

Such a system can be solved analytically using the \textit{exact invariant} method developed in \cite{lewis1968class,lewis1969exact} and summarized in \cite{pedrosa1997exact}. The general solution of equation \eqref{Schroedinger} is a linear combination $\chi_\nu = \sum_{n}c_\nu^{(n)}\chi_\nu^{(n)}$, were 
\begin{equation}\label{reducedQuantumSolution}
    \chi_\nu^{(n)} = \frac{H_n\Big(q_\nu/\sqrt{\hbar}\rho_\nu\Big)}{(\sqrt{\pi\hbar}n!2^n\rho_\nu)^{1/2}}\exp\Big[ i\varphi_{\nu}^{(n)} - \frac{q_\nu^2}{2\hbar\rho_\nu^2} \Big].
\end{equation}
Here, $H_n$ are the Hermite polynomials and $\varphi_{\nu}^{(n)}(\tau)$ are time dependent phases. We point out that, because in the following we are interested just in the calculus of the power spectra, the phases do not play any role and thus it is not necessary to calculate them. 

Therefore, the wave function of the perturbations is known once the functions $\rho_\nu(\tau)$ are given and these are the solutions of the differential equations
\begin{equation}
    \rho_\nu'' + \omega_\nu^2\rho_\nu = \rho_\nu^{-3}.
\label{GaussianWidthEquation}
\end{equation}

The general solution of Eq. \eqref{GaussianWidthEquation} can be obtained \cite{lewis1967classical} from the following relation
\begin{equation}\label{general gauss width solution}
    \rho_\nu = \frac{\gamma_1}{\alpha}\sqrt{ 2A^2g^2+2B^2f^2 +2\gamma_2\big[4A^2B^2-\alpha^2\big]^{1/2}fg },
\end{equation}
where $f(\tau)$ e $g(\tau)$ are two linearly independent solutions of the linear system associated to Eq. \eqref{GaussianWidthEquation}, 
$A$ and $B$ are arbitrary constants, $\alpha = fg'-gf'$ and $\gamma_{1,2} = \pm 1$.

As usual in the context of inflationary models we assume that the perturbations are initially in the ground state, which here correspond to $n=0$ for which the energy is minimal. Hence, we choose the following gaussian wave function
\begin{equation}\label{QuantumSolution}
    \chi_{\nu}(\tau,q_{\nu}) = \frac{1}{(\sqrt{\pi\hbar}\rho_\nu)^{1/2}}e^{-q_{\nu}^2/(2\hbar\rho_\nu^2)}.
\end{equation}
Now we must note that all the modes of astrophysical interest today had, at the beginning of inflation, a physical wavelength much smaller than the Hubble radius, which means that $k/aH \rightarrow\infty$. In this regime the perturbations are not affected by the expansion of the background and the modes behave as harmonic oscillators with constant frequencies $\omega_\textbf{k}=k$ and $\omega_T=K$. Hence, we will fix the constants $A$ and $B$ such that for each mode the wave function \eqref{QuantumSolution} approaches asymptotically the ground state of a minkowskian harmonic oscillator.

\subsection{\label{subsec:PowerSpectra}Power spectra of the perturbations}

Now that the wave function of the perturbations is known, we can derive the power spectra. 

Let us consider first the scalar sector. The quantum amplitudes $v_k$ have gaussian probability distributions. This is in agreement with the most recent measurements of the CMB anisotropies, which do not show any hint of primordial non-gaussianity \cite{planck2018_nongaussianity}. This implies that the statistical properties of the scalar modes are entirely given by the two-point correlation function 
\begin{equation}
\begin{split}
    \Xi_S = \braket{\chi|v_S(\textbf{x})v_S(\textbf{x}+\textbf{r})|\chi}.
\end{split}
\label{CorrFunc}
\end{equation} 
Here, $\ket{\chi} $ is the ground state of the scalar sector given in Eq. \eqref{WFansatz}. Making use of the Fourier transform of $v_S(\textbf{x})$ and the explicit expression of the wave functions \eqref{QuantumSolution}, the correlation function can be written as
\begin{equation}
    \Xi_S = \int\frac{d^3pd^3q}{(2\pi)^3}e^{i\textbf{p}\cdot\textbf{x}}e^{i\textbf{q}\cdot(\textbf{x}+\textbf{r})} \prod_{i,\textbf{k}'}\frac{1}{\sqrt{\pi\hbar\rho_{k'}^2}}I(\textbf{p},\textbf{q}),
\end{equation}
where we have defined the integral
\begin{equation}\label{integralPQ}
    I = \int \prod_\textbf{k}\Big(d{v^R_\textbf{k}}d{v^I_\textbf{k}}\ e^{-[{(v^R_\textbf{k})}^2+{(v^I_\textbf{k})}^2]/\hbar\rho_k^2}v_{\textbf{p}}v_{\textbf{q}}\Big).
\end{equation}
For $\textbf{p}\neq\pm\textbf{q}$, the integral \eqref{integralPQ} becomes linear in $v_\textbf{p}$ and $v_\textbf{q}$ and weighted by gaussian factors that make this contributions vanishing. For $\textbf{p}=\textbf{q}$ we have $v_{\textbf{p}}^2 = [(v^R_{\textbf{p}})^2 - (v^I_{\textbf{q}})^2 + 2iv^R_{\textbf{p}}v^I_{\textbf{q}}]/2$ so that the first two terms give the same contribution with opposite sign, while the third gives two linear integrals and the result again vanishes. Therefore, from the integration in $\textbf{p}$ and $\textbf{q}$ survive only the terms $\textbf{p}=-\textbf{q}$ for which $v_{\textbf{p}}v_{\textbf{-p}} = |v_{\textbf{p}}|^2 = [v_{R,\textbf{p}}^2 + v_{I,\textbf{q}}^2]/2$ and the final expression for the two-point correlation function is
\begin{equation}\label{corrFunction}
    \Xi_S = \int_0^\infty \frac{dp}{p} \frac{\sin pr}{pr}\Big(\frac{\hbar}{4\pi^2}p^3 \rho_p^2\Big).
\end{equation}
The power spectrum is defined as the Fourier amplitude of the variance $\Xi_S|_{r=0}=\braket{v_S^2}$ per unit logarithmic interval, thus it is given by
\begin{equation}\label{PSmukhanov}
    \mathcal P_v(k) = \frac{\hbar k^3}{4\pi^2}\rho_k^2.
\end{equation}

The connection between the quantum perturbations and the fluctuations of temperature and density appears in the comoving curvature perturbation $\zeta$. This variable has the feature of being conserved during all the period of expansion in which the perturbations lie outside the horizon \cite{Lyth1985Large-scale,weinberg2008cosmology}. Therefore, its spectrum, calculated at the end of inflation, can directly be propagated to the time when perturbations re-enter the horizon without the need to consider the details of the cosmological evolution. 

During the matter-dominated era, when the temperature anisotropies of the CMB are created, the relation of $\zeta$ to the Mukhanov-Sasaki variable is given by \cite{martin2012cosmological}
\begin{equation}\label{curvaturePerturation}
    \zeta_k = \sqrt{\frac{4\pi G}{\epsilon}}\frac{v_k}{a},
\end{equation}
where $\epsilon = 1 - \mathscr{H}'/\mathscr{H}^2$ is the slow-roll parameter. From Eqs. \eqref{PSmukhanov} and \eqref{curvaturePerturation} follows the expression of the power spectrum of the comoving curvature perturbation 
\begin{equation}\label{Scalar Pow Spec}
    \mathcal P_S(k) = \frac{\hbar G}{\pi\epsilon}k^3\frac{\rho_k^2}{a^2},
\end{equation}
where the ratio $\rho_k/a$ has to be calculated in the super-Hubble limit $\tau\rightarrow0$. 

\vspace{5mm}

For what concerns the tensor sector, the two-point correlation function factorizes into a time dependent quantum amplitude and a stochastic average containing the statistical properties of the spatial distribution. The latter is determined by the probability distribution governing the variable $\vartheta_{ij}$ and can be, in principle, different from the gaussian one, thus leading to non-gaussian features of the spectrum.  

It is easy to derive the time dependent component which is
\begin{equation}\label{TcorrFunc1} 
	\braket{\chi_T|v_T^2|\chi_T} = \int_{0}^\infty dv_T |\chi_T(\tau,v_T)|^2 v_T^2 =\frac{\hbar\rho_T^2}{4}.
\end{equation}
Assuming that the variable $\vartheta_{ij}$ is governed by a a translationally and rotationally invariant probability distribution function as well as space-inversion invariance, the stochastic term can be written as
\begin{equation}\label{XcorrFunc}
	\braket{\vartheta_{ij}(\textbf{x})\vartheta_{kl}(\textbf{x}+\textbf{r})} = \int \frac{d^3p}{(2\pi)^{3}} e^{-i\textbf{p}\cdot\textbf{r}}\Pi_{ij,kl}\mathcal{P}_\vartheta(p),
\end{equation}
where $\Pi_{ij,kl}(\hat p) = \sum_{\lambda}e_{ij}(\hat q,\lambda)e_{kl}^*(\hat q,\lambda)$, being $e_{ij}(\hat p,\lambda)$ the polarization tensor and $\lambda$ the helicity.

As for the comoving curvature perturbation also $\eta=\sqrt{4\kappa_g/\mathcal C_\vartheta}\,v_T/a$ becomes a constant, say $\bar\eta$, in the limit $\tau\rightarrow0$, i.e. when the perturbations are outside the horizon. This means that the tensor perturbation $\bar\eta\vartheta_{ij}$ provides an initial condition for the gravitational waves when they re-enter the horizon. From Eqs. \eqref{TcorrFunc1} and \eqref{XcorrFunc} we finally obtain the power spectrum of the tensor perturbations which reads as 
\begin{equation}\label{Tensor Pow Spec}
	\mathcal P_T(k) = \frac{16\hbar G}{\pi \mathcal{C}_\vartheta}\frac{\rho_T^2}{a^2}k^3\mathcal{P}_\vartheta(k),
\end{equation}
where, again, the ratio $\rho_T/a$ has to be evaluated in the super-Hubble limit.

We shall now apply what we have found here first to the de Sitter and then to the slow-rolling expansion.

\subsection{\label{subsec:deSitter}De Sitter case}

In the de Sitter case the potential of the scalar field takes a constant value $V (\phi) = \Lambda$ and plays the role of an effective cosmological constant. The unperturbed scalar field takes a constant value as well and, therefore, the gauge invariant scalar potential \eqref{ScalarGIvariable} reduces to the (gauge invariant) scalar field perturbation, $\xi_S = \varphi$. Note that the slow-roll parameter appearing in the power spectrum of the scalar modes \eqref{Scalar Pow Spec} should be zero. Indeed, in this scenario the matter-dominated era never happens, therefore the relation \eqref{curvaturePerturation} does not hold anymore and it makes no sense to speak about a power spectrum since the modes never re-enter the horizon. For this reason it is necessary to perform the computations imposing $\epsilon$ to be a non-vanishing constant.

The Hamilton-Jacobi equation \eqref{H-Jeq} reduces here to the following simple equation
\begin{equation}
    \Big(\frac{\partial S}{\partial a} \Big)^2=\frac{12}{\kappa_g}a^4\Lambda.
\end{equation}
From Eq. \eqref{velocities} it results that the explicit expression of the scale factor in terms of the new time coordinate reads as 
\begin{equation}\label{deSitterScaleFactor}
a(\tau) = -\frac{1}{H_0\tau},
\end{equation}
where $H_0 = \sqrt{\kappa_g\Lambda/3}$. Therefore $\tau$ is equal to the conformal time. 

It is now possible to express the frequencies as functions of time, which result to be the following
\begin{subequations}\label{eq:deSitterFreq}
\begin{align}
    &\omega_{k}^2 = k^2 - \frac{2}{\tau^2}, \label{Tfreq}\\[0.2cm]
    &\omega_T^2 = K^2 - \frac{2}{\tau^2} \label{Sfreq}.
\end{align}
\end{subequations}
The linear equation for the tensor gaussian width reads as $\rho_T''+(K^2-2/\tau^2)\rho_T=0$ and yields the following two linearly independent solutions
\begin{subequations}\label{linearSolution}
\begin{align}
    &f(\tau) = \frac{1}{K^{3/2}\tau}\big(K\tau\cos(K\tau) - \sin(K\tau)\big),\\[0.2cm]
    &g(\tau) = \frac{1}{K^{3/2}\tau}\big(\cos(K\tau) + K\tau\sin(K\tau)\big).
\end{align}
\end{subequations}
The general solution is thus given by inserting the functions \eqref{linearSolution} into Eq. \eqref{general gauss width solution}. The initial conditions discussed in Sec. \ref{subsec:Quantumdynamics} are satisfied by setting $A^2=B^2=1/2$ and $\gamma_1=1$ and give the following gaussian width 
\begin{equation}\label{TdeSitterGaussWidth}
    \rho_T(\tau) = \frac{\sqrt{1+K^2\tau^2}}{K^{3/2}\tau}.
\end{equation}
For what concerns the scalar modes, we note that the equation for $\rho_k$ is equal to the one for $\rho_T$ with the substitution $k\rightarrow K$. So it is sufficient to replace $K$ with $k$ in Eq. \eqref{TdeSitterGaussWidth} and we find the following scalar gaussian width
\begin{equation}\label{SdeSitterGaussWidth}
    \rho_\textbf{k}(\tau) = \frac{\sqrt{1+k^2\tau^2}}{k^{3/2}\tau}.
\end{equation}

We can finally give the expressions of the power spectra for the quasi-isotropic de Sitter evolution. This is done evaluating the ratios $\rho_k/a$ and $\rho_T/a$ in the super-Hubble limit and inserting the results into Eqs. \eqref{Scalar Pow Spec} and \eqref{Tensor Pow Spec}. We, thus, obtain the following power spectrum for the scalar modes
\begin{equation}\label{PSzeta}
\mathcal P_S(k) = \frac{\hbar GH_0^2}{\pi\epsilon}\Big\vert_{k=H_0a} ,
\end{equation}
where, because of the presence of the slow-roll parameter $\epsilon$, it has to be evaluated at the instant of time at which a certain mode-$k$ crosses the horizon, i.e. when $k = H_0a$; and the following power spectrum for the tensor modes
\begin{equation}\label{PST}
\mathcal P_T(k) = \frac{16
\hbar GH_0^2}{\pi \mathcal{C}_\vartheta K^3}k^3\mathcal{P}_\vartheta(k).
\end{equation}

Note that eq. \eqref{PSzeta} is the standard power spectrum of curvature perturbations for the de Sitter inflationary model, while Eq. \eqref{PST} results to be radically different from the standard scale invariant spectrum, since the scale dependence is contained in the function $\mathcal{P}_\vartheta(k)$ which is not constrained by the inflationary dynamics. In addiction, the tensor contribution is much smaller then the scalar one, being the tensor-to-scalar ratio $r = \frac{\mathcal P_T(k)}{\mathcal P_S(k)}$ proportional to $\epsilon$.

\subsection{\label{subsec:SlowRoll}Slow-rolling case}

Here we generalize the results \eqref{PSzeta} and \eqref{PST} to the regime in which the inflaton field undergoes a generic slow-roll dynamics. In this case the matter potential $V(\phi)$ develops a small dependence on $\phi$ and the background component of the scalar field acquires a finite velocity, i.e. $\phi'\neq 0$.

The analytical calculation of the power spectra is achieved making use of an expansion in the slow-roll parameters 
\begin{equation}\label{SRparameters}
    \epsilon = 1 - \frac{\mathscr{H}'}{\mathscr{H}^2}, \qquad \delta = 1 - \frac{\phi''}{\mathscr{H}\phi'},
\end{equation}
which must be small in order to guarantee the slow-roll approximation. Here we consider only the first order slow-roll approximation so that in the following we will drop quadratic and higher order terms.

First we need to derive the background quantities. Integrating the relation $d\tau=adt$, i.e.
\begin{equation}\label{conformal time 0}
    \tau = \int \frac{dt}{a} = \int \frac{da}{a^2H},
\end{equation}
we get the explicit expression of the scale factor in terms of the slow-roll parameters and conformal time
\begin{equation}\label{SR conformal time scale factor} 
    a(\tau) = \frac{C}{(-\tau)^{1+\epsilon}},
\end{equation}
where $C$ is an integration constant. Therefore, the Hubble parameter takes the following form
\begin{equation}\label{SR Hubbel parameter}
    H(\tau) = H_0\Big(\frac{\tau}{\tau_0}\Big)^\epsilon = H_0 \Big[1 + \epsilon\log\Big(\frac{\tau}{\tau_0}\Big)\Big],
\end{equation}
in which the constant $C$ it's been replaced by
\begin{equation}
    C = (1+\epsilon)\frac{(-\tau_0)^\epsilon}{H_0}.
\end{equation} 
Physically $H_0$ represents the reference de Sitter spacetime. We set it as the value of the Hubble parameter at the time $\tau_k$ of horizon crossing of the mode-$k$, that is $H_0 = H(\tau_k) = k/a(\tau_k)$, which implies $\tau_0(\tau_k) = -1/k$. Analogously, for what concerns the tensor degree of freedom we choose to set $H_0=K/a$ and $\tau_0=1/K$, since $v_T$ behaves like an oscillator with comoving wave-vector norm $K$. In this way both $H_0$ and $\tau_0$ become $k$-dependent.

Note that eq. \eqref{SR Hubbel parameter} fixes the range of validity of the approximation, that is given by the request that the expansion rate is almost constant. Thus we must impose $\epsilon|\log(\tau/\tau_0)|\ll 1$, that gives
\begin{equation}
    e^{-1/\epsilon} \ll \frac{\tau}{\tau_0} \ll e^{1/\epsilon},
\end{equation}
and assuming $\epsilon$ to be very small, this range is very large.

The Friedmann equation can be written without approximations in terms of the first slow-roll parameter as 
\begin{eqnarray}
    V = \frac{1}{\kappa_g}(3-\epsilon)H^2.
\end{eqnarray}
It is easy then to derive the explicit expressions of $V$ and $\phi'$ in term of $\tau$ and finally find the frequencies \eqref{scalarfreq} and \eqref{tensorfreq}, which results to be the following
\begin{subequations}
\begin{align}
    {\omega_{k}}^2 = k^2 - \frac{2+3\gamma}{\tau^2} \label{SR scalar frequency},\\[0.2cm]
    {\omega}_T^2 = K^2 - \frac{2+3\epsilon}{\tau^2},\label{SR tensor frequency}
\end{align}
\end{subequations}
where we have defined the parameter $\gamma = 2\epsilon-\delta$. 

The linearly independent solutions $f(\tau)$ and $g(\tau)$  of the linear system of differential equations for the gaussian widths can be obtained in terms of the Bessel functions $J_{\pm(3/2+\gamma)}(-k\tau)$ and $J_{\pm(3/2+\epsilon)}(-K\tau)$, for the scalar and tensor sector respectively. Again, the gaussian widths are constructed from Eq. \eqref{general gauss width solution} and imposing the ``minkowskian" initial conditions. In the super-Hubble limit they read as
\begin{subequations}\label{SRgauss}
\begin{align}
    &\rho_k = \frac{\Gamma(3/2+\gamma)}{2^{-(1+\gamma)}\sqrt{\pi k}}(-k\tau)^{-(1+\gamma)}, \label{SRgauss1}\\[0.3cm]
    &\rho_T = \frac{\Gamma(3/2+\epsilon)}{2^{-(1+\epsilon)}\sqrt{\pi K}}(-K\tau)^{-(1+\epsilon)}. \label{SRgauss2} 
\end{align}
\end{subequations}

Now we can derive the power spectra linearizing Eq. \eqref{SRgauss} with respect to the slow-roll parameters and evaluating the ratio between the gaussian widths and $a$ in the super-Hubble limit. In this way Eq. \eqref{Scalar Pow Spec} gives the following power spectrum of the curvature perturbation 
\begin{equation}\label{SRPSscalar}
\mathcal P_\zeta(k) = [1-2\epsilon + 2\gamma(2-\gamma_E-\log 2)]\frac{\hbar GH_0^2}{\pi\epsilon}\Big\vert_{aH=k},
\end{equation} 
and eq. \eqref{Tensor Pow Spec} gives the power spectrum of the tensor perturbations which reads as
\begin{equation}\label{SRPStensor}
\begin{split}
    \mathcal P_T(k) = &[1 + 2\epsilon(1-\gamma_E-\log 2)]\\
    &\times\frac{16\hbar GH_0^2}{\mathcal C_\vartheta K^3}\Big\vert_{aH=k}k^3\mathcal{P}_\vartheta(k).
\end{split}
\end{equation}
Again, the result \eqref{SRPSscalar} is the standard power spectrum of the scalar perturbations while in \eqref{SRPStensor} the contribution of the mode-$k$ is arbitrary, i.e. inflation determines only the overall constant factor not the contribution of each mode to the total variance. Furthermore, the tensor-to-scalar ratio is proportional to $\epsilon \ll 1$, therefore the tensor contribution is much smaller than the scalar one.

\section{Concluding remarks}\label{sec:Conclusions}

Our analysis was based on the implementation of the so-called quasi-isotropic solution in the early Universe dynamics, in which the standard FLRW Universe presents a small 
inhomogeneous correction, evolving according to a second scale factor different from the background one. The smallness of the inhomogeneities is guaranteed by the correspondingly small ratio $\eta(t)$ between the FLRW scale factor and the perturbation one, which has been taken as the fundamental dynamical variable in our study. 

Here, we implemented the idea proposed in \cite{vilenkin1989interpretation} of considering the inhomogeneous correction and, in particular, the small variable controlling their values, as associated to a ``small quantum" subspace, living on the quasi-classical framework of the isotropic Universe. 

In this way, the variable $\eta(t)$ was, \textit{de facto}, treated as a small quantum correction to the standard primordial 
cosmology, whose dynamics is described by a Schr\"odinger equation, resembling a time dependent harmonic oscillator. 

We studied the resulting dynamics of the proposed picture during the de Sitter phase of an inflationary era and we involved in the problem of characterizing both the scalar perturbation of the corresponding inflaton field and the tensor perturbation of the gravitational degrees of freedom. 

We constructed the Lagrangian and Hamiltonian formulation of this cosmological scenario, up to the second order of approximation in the perturbation amplitude and adopting the gauge invariant formulation, guaranteed by the use of the Mukhanov-Sasaki variables. 

We then calculated the spectrum of the primordial fluctuations emerging from the de Sitter phase, comparing the tensor and scalar spectrum properties. While the latter retains its standard scale invariant form, the former has a peculiar dependence on the wave-vectors, fixed, \textit{ab initio}, from the nature of the considered inhomogeneous term perturbing the isotropic Universe.

Actually, the resulting feature of our dynamical setting is that the tensor spectrum, emerging from inflation, can have a generic (non-necessarily gaussian) profile.

The obtained result is a direct consequence of the factorization of the time and space dependence that naturally takes place in the quasi-isotropic solution. Indeed, we have shown that, while the scalar component behaves as an intrinsically inhomogeneous degree of freedom, such a quasi-isotropic character leads to a factorization of the tensor component of the metric. Thus, if this feature can be postulated also for a quantum Universe (no objection exists \emph{a priori} to this statement), we can infer that, since the isotropic Universe emerges as the main classical component of the dynamics, a factorized perturbation can still evolve on such a background, becoming responsible for a generic spectral feature of the primordial tensor spectrum. 

The robustness of the quasi-classical solution in the Planckian representation of the Universe could also be inferred from a quantum dynamical process of isotropization of the generic cosmological solution \cite{belinskii1982general,montani1995general,kirillov1993nature}. According to the ideas discussed in \cite{kirillov1997origin}, the emergence of a classical Universe from a quantum Mixmaster scenario \cite{benini2006inhomogeneous,montani2008classical} can approach a classical limit only when the anisotropy degrees of freedom becomes small enough and a quasi-classical isotropic Universe starts to emerge. It is exactly this dynamical picture the most convincing motivation for the present analysis.

Finally, we point out that at a classical level it was demonstrated in \cite{lukash1974bianchiI,lukash1973BianchiIX,lukash1976quantumeffects} that the backreaction on the metric of the particles produced during the BKL oscillatory regime leads to a quick isotropization of the cosmological expansion, actually giving a quasi-isotropic evolution of the Universe. Therefore, it would be interesting to see if this mechanism holds also in a quantum regime and in the presence of a scalar field, thus allowing one to study the space distribution of the tensor inhomogeneity $\vartheta_{ij}(\textbf{x})$ and linking the tensor power spectrum to the pre-inflationary Universe.

\bibliography{Bibliography}

\end{document}